\documentclass[prl,showpacs,twocolumn,amsmath,amssymb,groupadress,superscriptaddress]{revtex4}
\usepackage{graphicx}% Include figure files

\begin{document}

\title{Electric field switching of antiferromagnetic domains in YMn$_2$O$_5$: a probe of the multiferroic mechanism}

\author{P.G. Radaelli}
\affiliation{ISIS facility, Rutherford Appleton Laboratory-STFC,
Chilton, Didcot, Oxfordshire, OX11 0QX, United Kingdom. }
\affiliation{Dept. of Physics and Astronomy, University College
London, Gower Street, London WC1E 6BT, United Kingdom}
\author{L.C. Chapon}
\affiliation{ISIS facility, Rutherford Appleton Laboratory-STFC,
Chilton, Didcot, Oxfordshire, OX11 0QX, United Kingdom. }
\author{A. Daoud-Aladine}
\affiliation{ISIS facility, Rutherford Appleton Laboratory-STFC,
Chilton, Didcot, Oxfordshire, OX11 0QX, United Kingdom. }
\author{C. Vecchini}
\affiliation{ISIS facility, Rutherford Appleton Laboratory-STFC,
Chilton, Didcot, Oxfordshire, OX11 0QX, United Kingdom. }
\affiliation{Institute of Electronic Structure and Laser, Foundation for Research and Technology - Hellas, Vassilika Vouton, 711 10 Heraklion, Crete, Greece. }
\author{P.J. Brown}
\affiliation{Institut Laue-Langevin, 6, rue Jules Horowitz, BP 156 - 38042 Grenoble Cedex 9 - France.}
\author{T. Chatterji}
\affiliation{Institut Laue-Langevin, 6, rue Jules Horowitz, BP 156 - 38042 Grenoble Cedex 9 - France.}
\author{S. Park}
\affiliation{Department of Physics and Astronomy, Rutgers
University, Piscataway, New Jersey 08854, USA}
\author{S-W. Cheong}
\affiliation{Department of Physics and Astronomy, Rutgers
University, Piscataway, New Jersey 08854, USA}
\date{\today}% It is always \today, today,

\begin{abstract}
We employ neutron spherical polarimetry to determine the nature and population of the coexisting antiferromagnetic domains in multiferroic YMn$_2$O$_5$.  By applying an electric field, we prove that reversing the electrical polarization results in the population inversion of two types of in-plane domains, related to each other by inversion.  Our results are completely consistent with the exchange striction mechanism of ferroelectricity, and support a unified model where cycloidal ordering is induced by coupling to the main magnetic order parameter.
\end{abstract}


\maketitle

For a long time, ferroelectricity and magnetism have been thought of as essentially independent. Recently however, a class of "novel" multiferroics, such as TbMnO$_3$ \cite{Kimura_Nature03} and \textit{RE}Mn$_2$O$_5$ (\textit{RE}=Y, rare earth, Bi) \cite{Hur_Nature04} has been discovered, in which antiferromagnetic ordering itself is though to induce ferroelectricity.  Unlike 'conventional' multiferroics, magneto-electric coupling in these materials is very strong, leading to spectacular effects, such as rotation or reversal of the electrical polarization upon application of a magnetic field \cite{Cheong_NatureMat}.  The mechanism (or mechanisms) leading to ferroelectricity upon magnetic ordering are not known with certainty \cite{Khomskii_JMMM2006}.  Most theories \cite{Harris_Lawes_condmat,Mostovoy_PRL_06} have been guided by the observation that the majority of these materials possess a cycloidal structure, in which the spin direction is rotated continuously, and have postulated that ferroelectricity arises from displacements, due to spin-orbit coupling, of ionic or electronic density between two metal ions whose spins are non-collinear.  In its high-temperature commensurate phase, The YMn$_2$O$_5$ magnetic structure is almost collinear \cite{ChaponPRL_Y,Noda_PhysB06} -  its spins are only slightly tilted ( 15$^{\circ}$) away from collinearity -  but it has been shown \cite{Noda_PhysB06} that this small tilt, including a weak cycloidal component in the crystallographic $bc$ plane, could in principle induce ferroelectric displacements through spin-orbit coupling.  An alternative mechanism, not requiring non-collinearity, has been proposed \cite{ChaponPRL_Y} for YMn$_2$O$_5$, in which small displacements of oxygen or metal ions optimize magnetic superexchange energy (exchange striction).  Regardless of the specific mechanism, a consistent prediction for these 'novel' multiferroics is that different directions of the electrical polarization should be associated with distinct antiferromagnetic domains.  In general, one would expect these domains to be related by the symmetry operators that are lost at the ferroelectric  transition, namely the center of inversion.  However, the case of \textit{RE}Mn$_2$O$_5$ is not trivial in this respect: if only one if the two components (cycloidal or planar) is coupled to the electrical polarization, it is conceivable that only that component will respond to an applied electric field.  This possibility is particularly appealing in the case of the spin-orbit mechanism, since in this case the polarization could be reversed by switching the small $c$-axis component $S_z$, while leaving the large in-plane component $S_{xy}$, and consequently the main antiferromagnetic domain structure, unaltered.  Whether this will happen depends on the strength of the magnetic coupling between the two components, which, however, can be very weak, since the only coupling terms allowed are antisymmetric \cite{Note_0}. Observation of these domains and of the domain switching by an external electric field provides therefore a powerful insight in the coupling of the two substructures to the polarization and to each other. Optical second-harmonic generation (SHG) \cite{Fiebig_Nature_02,Lottermoser_Nature_04} and scattering of polarized neutrons \cite{Yamasaki_PRL07} have been previously employed to observe domain formation and switching in multiferroics.  However, SHG is intrinsically limited to magnetic ordering at the gamma point, and scattering of polarised neutrons (without analysis) affords limited information about the structure of the domains, which must be known in advance.  In this Letter, we describe a direct observation of the YMn$_2$O$_5$ antiferromagnetic domains using neutron spherical polarimetry, which is uniquely sensitive to both domain structure and domain population.  By cooling the crystal in positive or negative external electric fields,  we determine the in-plane magnetic structures of the two domain types uniquely, and conclude that switching from one to the other involves reversal of the in-plane components $S_{xy}$ of the spins, not just a change in the small $S_z$ component. The results of this experiment are most easily understood in the framework of the exchange striction model, where $S_{xy}$ is directly coupled to the polarization, and also suggest that $S_z$ is induced by the \emph{direct} Dzyaloshinskii-Moriya (DM) interaction.

\begin{figure}[!h]
\includegraphics[scale=0.5]{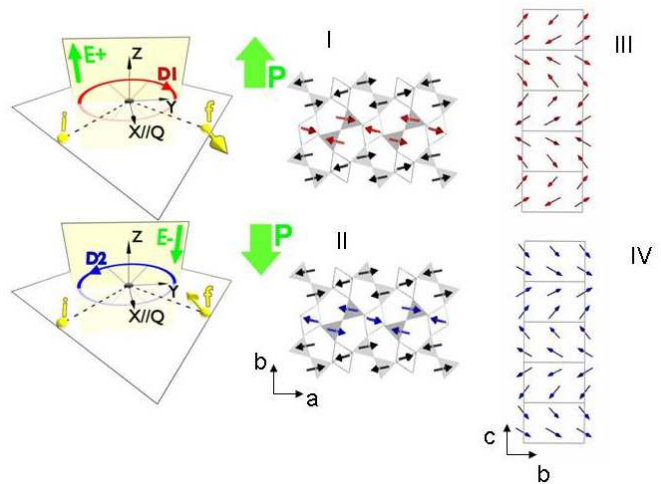}
\caption{(color online) \textbf{left}:  schematic representation of the neutron spherical polarimetry experiment for the two domains, here shown in the idealized case of an unpolarized incident beam.  The real (or imaginary) component of the magnetic structure factor projection $\textbf{\textrm{M}}^{\bot}_{hkl}$ rotate clockwise (counter clockwise) for Domain \textbf{I} (Domain \textbf{II}), creating a spin polarization of opposite signs for the scattered neutron.  The direction of the applied electric field is also indicated.  right: Magnetic structures of YMn$_2$O$_5$ for the different domain configurations, projected on the $ab$ plane (\textbf{I} and \textbf{II}) and the $bc$ plane (\textbf{III} and \textbf{IV}).  Small arrows represent magnetic moments.   The observed domain switching mechanism is represented by the inversion (change from the red to the blue) in the central chain (between configurations \textbf{I} and \textbf{II} in the $ab$  plane). \label{Fig: polarimetry}}
\end{figure}

The YMn$_2$O$_5$ magnetic structure for 24 K $<$ T $<$ 38 K has been determined from neutron powder \cite{ChaponPRL_Y} and single crystal \cite{Noda_PhysB06,vecchini} data, and is shown in  Fig \ref{Fig: polarimetry} for the possible polar domain configurations.  It consists of staggered antiferromagnetic zig-zag chains running along the crystallographic $a$ axis (horizontal in the figure, panels \textbf{I} and \textbf{II}), with spins parallel within each chain and tilted by about 15$^{\circ}$  with respect to the $a$  axis.  Configurations \textbf{I} and \textbf{II} are related by inversion, although they can also be obtained by reversing the direction of the spins in half of the chains \cite{vecchini}. The sign of the out-of-phase $c$-axis component (configurations \textbf{III} and \textbf{IV} also related by inversion) determines the rotation direction of the cycloidal modulation in the $bc$-plane.
Unique domains are obtained by combining in-plane and cycloidal components: the combination \textbf{I}+\textbf{III} is related by overall inversion symmetry to \textbf{II}+\textbf{IV} and so is \textbf{I}+\textbf{IV} with \textbf{II}+\textbf{III}. Domains differing solely by the sign of the c-axis component, like \textbf{I}+\textbf{III} and \textbf{I}+\textbf{IV} are not related by symmetry and are in principle distinguishable by diffraction. Indeed, neutron diffraction data clearly favor the \textbf{I}+\textbf{III} and \textbf{II}+\textbf{IV} configurations.  However, an admixture with the other domains would result in a reduction of the refined $S_z$ component, and can not be completely ruled out. As already mentioned, the energetic degeneracy between these domains is lifted only by weak antisymmetric exchange \cite{Note_0}.

The single crystals of YMn$_2$O$_5$ was grown using B$_2$O$_3$/PbO/Pb flux in a Pt crucible. The flux was held at 1,280 $^{\circ}$C for 15 hours and slowly cooled down to 950 $^{\circ}$C at a rate of 1 $^{\circ}$C per hour. Crystals grew in the form of cubes. The sample dimension was 4x0.8x4 mm$^{3}$ along the crystallographic $a$, $b$ and $c$ directions.  Thin silver-paint electrodes were attached to the crystal faces perpendicular to the b-axis, with gold wires providing electrical connections to the voltage supply.  The sample was attached to an Al support with "GE" varnish, and mounted with the b-axis perpendicular to the scattering plane. Neutron spherical polarimetry data were collected  as a function of temperature using the diffractometer D3 at the Institut Laue-Langevin (Grenoble, France) equipped with CRYOPAD-II.   The crystal was mounted with the b-axis (i.e., the direction of the electrical polarization) perpendicular to the diffraction plane (vertical), so that only $h0l$-type Bragg peaks were measured.  An external electric field of up to 2.2  kV/cm and of either polarity was applied to the crystal along the $b$ axis.  Neutron spherical polarimetry is described in detail elsewhere \cite{Tasset_99} and also summarized in the EPAPS supplementary material.  Briefly, we can align the spins of the incident neutrons in any chosen direction and determine both the magnitude and direction of the polarisation of the scattered beam.

The polarization of the scattered beam for each Bragg peak $hkl$ can be calculated for any incident polarization from the nuclear structure factor $N_{hkl}$  and the magnetic structure factor $\textbf{\textrm{M}}_{hkl}$ of the crystal \cite{Blume_PR_63,Maleev_SPSS_63}.  Like the familiar X-ray structure factor, $N_{hkl}$  is a complex number, whereas $\textbf{\textrm{M}}_{hkl}$ is a complex vector.  Neutron diffraction and polarimetry are only sensitive to the projection of $\textbf{\textrm{M}}_{hkl}$ \emph{perpendicular} to the scattering vector $\textbf{\textrm{Q}}_{hkl}$, here indicated as $\textbf{\textrm{M}}^{\bot}_{hkl}$.  The equations for neutron polarimetry are greatly simplified when nuclear and magnetic scattering do not interfere and when the incident beam is fully polarized, as in the present case \cite{Note_2}.  These equations are best expressed in the so-called Blume reference frame \cite{Blume_PR_63}, with the X-axis parallel to the scattering vector (i.e., bisecting the incident and scattered beam), the Z-axis perpendicular to the scattering plane (vertical in our case) and the Y-axis completing the right-handed set (see Fig \ref{Fig: polarimetry}).   Noting that $\textbf{\textrm{M}}^{\bot}_{hkl}$ lies in the YZ plane, without loss of generality we can write:

\begin{equation}
\label{Eq: ellipse}
\textbf{\textrm{M}}^{\bot}_{hkl} =\left| \textbf{\textrm{M}}^{\bot}_{hkl} \right|\,e^{i\psi}\left(\hat{\textbf{y}}\cos \alpha+\hat{\textbf{z}}e^{i\phi} \sin \alpha\right)
\end{equation}

Where $\hat{\textbf{y}}$ and $\hat{\textbf{z}}$ are unit vectors in the Y and Z directions.  If the incident beam is fully polarized, the equations are further simplified, and can be expressed using the matrix elements $P_{ij}$ ($i,j$=X, Y , Z), which represent the polarization measured in the direction $j$ if the incident beam is polarized in the direction $i$

\begin{eqnarray}
P_{xx}&=& 1 \nonumber\\
P_{yy}=-P_{zz}&=& \cos 2 \alpha\nonumber\\
P_{yx}=P_{zx}&=& \sin 2 \alpha \, \sin \phi\nonumber\\
P_{zy}=P_{yz}&=& \sin 2 \alpha \, \cos \phi\nonumber\\
P_{xy}=P_{xz}&=&0
\end{eqnarray}

These matrix elements result from two terms:  the rotated polarization, which is proportional to the incident polarization, and the created polarization (elements $P_{yx}$ and $P_{zx}$), which is present even if the incident beam is unpolarized, and is always parallel to the scattering vector (i.e., to X).  The best way to visualize Eq. \ref{Eq: ellipse} is the following: for all but the simplest magnetic structures, the phase factor $\psi$  in \ref{Eq: ellipse} varies from one chemical unit cell to the next along the so-called "propagation" direction, so that the real and imaginary components of $\textbf{\textrm{M}}^{\bot}_{hkl}$ describe an ellipse in the YZ plane (Fig \ref{Fig: polarimetry}).  The parameters $\alpha$ and $\phi$  define the cardinal equation of the ellipse, while the sign of $\phi$  establishes whether the rotation around the ellipse is clockwise or counter clockwise.  The unique sensitivity of spherical polarimetry to the domain structure stems from the fact that for the same reflection $hkl$ the rotation is opposite for inversion domains, and consequently the signs of the created polarization $P_{yx}$ and $P_{zx}$ are reversed (Fig. \ref{Fig: polarimetry}).  Crucially, the magnetic structure itself need not be chiral or even rotating, but cannot be perfectly collinear, because in this case the created polarization will vanish.

In the geometry we employed, the polarization matrix elements are essentially insensitive to the small $c$-axis  spin component \cite{Noda_PhysB06,vecchini}, so we only probe domain configurations \textbf{I} and \textbf{II}, which, as already mentioned, are related by inversion.  The parameters   and   for this structure are readily calculated for each Bragg reflection $hkl$.  In particular, for Domain \textbf{I}   $\phi\approx +90^{\circ}$ for $h=2n+ \frac{1}{2}$,  $\phi\approx -90^{\circ}$  for $h=(2n+ 1)+\frac{1}{2}$, while the signs are reversed for Domain \textbf{II}.  The parameter $\alpha$  depends on $hkl$, but does not change between the two domains for a given Bragg peak.

\begin{figure}[h!]
\includegraphics[scale=0.50]{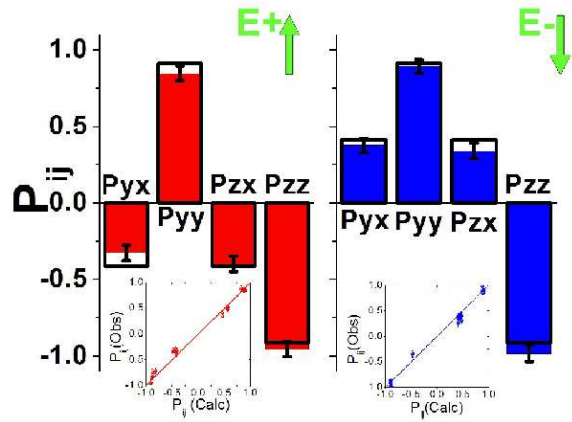}
\caption{(\textbf{main}:  observed (solid/color and error bars) and calculated (rectangles) neutron spherical polarimetry matrix elements at 25 K for the $-\frac{1}{2}\, 0\, -\frac{5}{4}$  Bragg peak of an YMn$_2$O$_5$ crystal cooled in a positive (left, red color) and negative (right, blue color) electric field of $\pm$ 2.2 kV/cm.  Note how the signs of the created neutron polarization elements $P_{yx}$ and $P_{zx}$ are reversed between the two field orientation, while all the absolute values and the signs of the other elements stay the same - a clear indication of domain population reversal.  \textbf{insets} observed and calculated matrix elements for several Bragg peaks in the two field orientations.  The model has no adjustable parameter, and assumes 100 \% population of one of the two domains.(color online).\label{Fig: 25K_fit}}
\end{figure}

The results of our spherical polarimetry measurements at 25 K are displayed in Fig. \ref{Fig: 25K_fit}.  The main panel shows the significant matrix elements for the reflection $-\frac{1}{2} \, 0 \,-\frac{5}{4}$, while the inset summarizes all results for several Bragg peaks.   Data on the left and right panels were collected after cooling the crystal through the magnetic transition (T$_N$=40 K) under a positive or negative electric field, respectively (E= $\pm$ 2.2 kV/cm).  The calculated values, based on the previously described magnetic structure and either 100 \% population of Domain \textbf{I} or Domain \textbf{II} (no adjustable parameters), are in excellent agreement with the observations.  In particular, the sign of the created polarisation ($P_{yx} =P_{zx}$) is reversed between the two electric field polarities, whereas the magnitudes and the signs of the other elements are unchanged, as expected for a complete population reversal.

\begin{figure}[h!]
\includegraphics[scale=2.3]{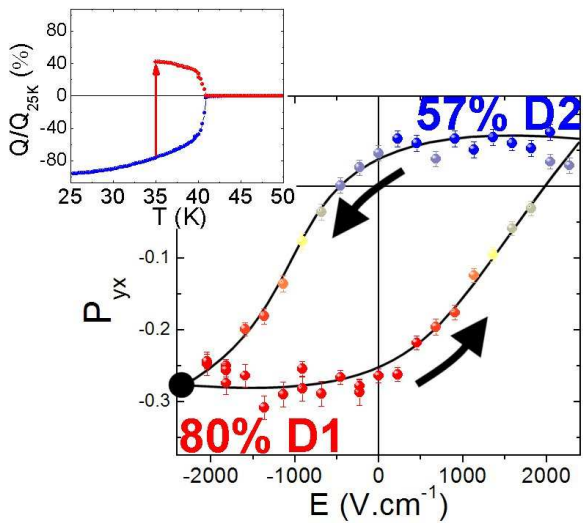}
\caption{ \textbf{main}:  partial hysteresis loop measured on the created neutron polarization element $P_{yx}$  for the $-\frac{1}{2}\, 0\, -\frac{7}{4}$  Bragg peak of an YMn$_2$O$_5$ crystal, warmed to 35 K after previous cooling to 25 under a negative -2.2 kV/cm electric field.  \textbf{inset}:  integrated pyroelectric currents measured on a 0.5 mm thick YMn$_2$O$_5$ crystal of the same batch on cooling to 25 K in a negative -2.0 kV/cm electric field (bottom/blue curve), followed by warming to 35 K and switching to a positive +2.0 kV/cm electric field (top/red curve).  The data are normalized to the fully saturated value at 25 K.  Both hysteresis bias and the switching ratios are in very good agreement with the neutron data. (color online) \label{Fig: histeresis}}
\end{figure}

At 25 K, the domains are strongly pinned, and cannot be switched by reversing the electric field at constant temperature.  Therefore, we warmed the crystal to 35 K, which is sufficiently close to the transition to observe field switching.  The domain population was still biased by the previous field cooling, so we could only observe a partial hysteresis loop (Fig. \ref{Fig: histeresis}).  Here, the sign of the created neutron polarization is reversed, but the magnitudes are not the same for +2.2 kV/cm and -2.2 kV/cm, indicative of domain ratios 80 \%-20 \% and 43 \%-57 \%, respectively.  In a separate experiment, the electrical polarization of a crystal of the same batch was measured under similar conditions.  Both the hysteresis bias and the switching ratios were found to be in very good agreement with our neutron data (inset to Fig \ref{Fig: histeresis}.)

The simplest version of the spin-orbit model, directly derived from TbMnO$_3$ and related materials, combined with the observation that the magnetic coupling between cycloids and planar structure can be weak, naturally leads to the prediction that only $S_z$  should reverse with the electric field --- a prediction that is strongly contradicted by our data.  Conversely, the reversal of the in-plane spin components is a strong prediction of the exchange striction mechanism \cite{ChaponPRL_Tb}, and is here completely verified.  We remark, however, that our data are completely consistent with a electric-field-driven transition between inversion-related domains in which both the  $S_{xy}$ and $S_z$ components switch in alternate chains. In this scenario, which we believe to be the most plausible, the in-plane and cycloidal components would always have the same polarity for a given direction of the polarization, and could in principle both contribute to ferroelectricity.  The relative importance of the two effects can however be gauged based on the same neutron data \cite{vecchini}:  for YMn$_2$O$_5$, the cross product $\textrm{\textbf{S}}_i  \times \textrm{\textbf{S}}_j$ (related to the spin-orbit mechanism) is about 40 times smaller than the dot product $\textrm{\textbf{S}}_i  \cdot \textrm{\textbf{S}}_j$ and a factor of 100 smaller than $\textrm{\textbf{S}}_i  \times \textrm{\textbf{S}}_j$ for multiferroic TbMnO$_3$ which displays a spontaneous polarization similar to that observed for YMn$_2$O$_5$.  Therefore, switching of the $S_z$ component, if observed, would indicate that antisymmetric coupling between cycloids and planar structure is important for the overall stability of the magnetic structure, strongly suggesting that the spirals are in fact stabilized by \emph{direct} Dzyaloshinskii-Moriya (DM) rather than next-nearest-neighbor interaction.  A unified picture would therefore emerge, in which the frustrated geometry of the Mn ions is primarily responsible for breaking inversion symmetry upon magnetic ordering, while both cycloidal ordering and ferroelectricity are induced by coupling to the main polar order parameter.

%\bibliography{RMn2O5_Manuscript}

\begin{thebibliography}{18}
\bibitem[{Not({\natexlab{b}})}]{Note_2}
\bibinfo{note}{The magnetic propagation vector is $\frac{1}{2} \,0
  \,\frac{1}{4}$ , so magnetic and nuclear Bragg peaks never coincide.}
\end{thebibliography}

\end{document}